\documentclass[conference]{IEEEtran}
\IEEEoverridecommandlockouts
\usepackage{amsmath,amssymb,amsfonts}
\usepackage{graphicx}
\usepackage{textcomp}
\usepackage{xcolor}
\def\BibTeX{{\rm B\kern-.05em{\sc i\kern-.025em b}\kern-.08em
    T\kern-.1667em\lower.7ex\hbox{E}\kern-.125emX}}

\usepackage{comment}

\usepackage{nohyperref}

\definecolor{matplotlib0}{HTML}{1f77b4}
\definecolor{matplotlib1}{HTML}{d62728}
\definecolor{matplotlib2}{HTML}{2ca02c}
\definecolor{matplotlib3}{HTML}{ff7f0e}
\definecolor{matplotlib4}{HTML}{9467bd}
\definecolor{matplotlib5}{HTML}{8c564b}
\definecolor{matplotlib6}{HTML}{e377c2}
\definecolor{matplotlib7}{HTML}{7f7f7f}
\definecolor{matplotlib8}{HTML}{bcbd22}
\definecolor{matplotlib9}{HTML}{17becf}

\usepackage{mathtools} % \abs, ...

\usepackage{booktabs}
\usepackage{multirow}
\usepackage{colortbl}
\usepackage{tablefootnote}
\usepackage{threeparttable}

\usepackage[acronym, style=super, nonumberlist]{glossaries}

\newcommand{\wolf}[0]{{Mr.\,Wolf}}
\newcommand{\riscy}[0]{\textsc{RI5CY}}
\newcommand{\eegnet}[0]{\textsc{EEGNet}}
\newcommand{\pulp}[0]{\textsc{PULP}}

\newcommand{\ibex}[0]{\textsc{Ibex}}
\newcommand{\riscv}[0]{\textsc{RISC-V}}
\newcommand{\xpulp}[0]{\textsc{XpulpV2}}

\usepackage{tikz}
\usetikzlibrary{shapes, arrows, snakes}
\tikzset{>=latex}

\usepackage{setspace} % used to set line spacing inside tikz figure

\tikzstyle{block} = [draw, thick, rectangle, minimum height=0.75cm, minimum width=0.75cm]
\tikzstyle{sum} = [draw, fill=white, circle, node distance=1cm, thick]
\tikzstyle{gain} = [
  regular polygon,
  regular polygon sides=3,
  draw,
  thick,
  fill=white,
  text width=0.3em,
  inner sep=0.3mm,
  outer sep=0mm,
  shape border rotate=-90
]
\tikzstyle{revgain} = [
  regular polygon,
  regular polygon sides=3,
  regular polygon rotate=90,
  draw,
  thick,
  fill=white,
  text width=0.3em,
  inner sep=0.3mm,
  outer sep=0mm,
]

\pgfdeclaresnake{eeg}{initial} {
  \state{initial}[width=10pt, next state=low] {
    \pgfpathcurveto{\pgfpoint{5pt}{8pt}}{\pgfpoint{5pt}{-8pt}}{\pgfpoint{10pt}{0pt}}
  }
  \state{low}[width=10pt, next state=initial] {
    \pgfpathcurveto{\pgfpoint{5pt}{4pt}}{\pgfpoint{5pt}{-4pt}}{\pgfpoint{10pt}{0pt}}
  }
  \state{final} {
    \pgfpathlineto{\pgfpoint{\pgfsnakeremainingdistance}{0pt}}
  }
}

\pgfdeclaresnake{eeg filter1}{initial} {
  \state{initial}[width=12pt, next state=low] {
    \pgfpathcurveto{\pgfpoint{6pt}{8pt}}{\pgfpoint{6pt}{-8pt}}{\pgfpoint{12pt}{0pt}}
  }
  \state{low}[width=10pt, next state=initial] {
    \pgfpathcurveto{\pgfpoint{6pt}{4pt}}{\pgfpoint{6pt}{-4pt}}{\pgfpoint{12pt}{0pt}}
  }
  \state{final} {
    \pgfpathlineto{\pgfpoint{\pgfsnakeremainingdistance}{0pt}}
  }
}

\pgfdeclaresnake{eeg filter2}{initial} {
  \state{initial}[width=8pt, next state=low] {
    \pgfpathcurveto{\pgfpoint{4pt}{8pt}}{\pgfpoint{4pt}{-8pt}}{\pgfpoint{8pt}{0pt}}
  }
  \state{low}[width=10pt, next state=initial] {
    \pgfpathcurveto{\pgfpoint{4pt}{4pt}}{\pgfpoint{4pt}{-4pt}}{\pgfpoint{8pt}{0pt}}
  }
  \state{final} {
    \pgfpathlineto{\pgfpoint{\pgfsnakeremainingdistance}{0pt}}
  }
}

\usepackage{pgfplots}
\usepgfplotslibrary{fillbetween}
\usepgfplotslibrary{colormaps}
\pgfplotsset{compat=1.16}

\pgfplotscreateplotcyclelist{matplotlib}{
  {matplotlib0},
  {matplotlib1},
  {matplotlib2},
  {matplotlib3},
  {matplotlib4},
  {matplotlib5},
  {matplotlib6},
  {matplotlib7},
  {matplotlib8},
  {matplotlib9}
}

\pgfplotsset{every axis/.append style={
    cycle list name=matplotlib,
}}

\usepackage{listings}

\definecolor{code_default}{HTML}{000000}
\definecolor{code_keyword}{HTML}{AC4142}
\definecolor{code_identifier}{HTML}{D28445}

\lstdefinelanguage{RISCV}{
  sensitive=false,
  morecomment=[l]{//},
  alsoletter={.},
  morekeywords=[1]{
    lp.setup, mv, lw, p.lw, sw, p.sw, pv.sdotsp.b, pv.shuffle2.b, p.subNR, p.addNR
  },
  morekeywords=[2]{
    zero, ra, sp, gp, tp, t0, t1, t2, t3, t4, t5, t6, s0, s1, a0, a1, a2, a3, a4, a5, a6, a7, a8, a9, a10, a11,
  },
  morestring=[b]",
  morestring=[b]',
}[strings, comments, keywords]

\lstdefinestyle{RISCV_STYLE}{
  language=RISCV,
  numbers=none,
  basicstyle=\scriptsize\ttfamily\color{code_default},
  keywordstyle=[1]\color{matplotlib0},
  keywordstyle=[2]\color{matplotlib1},
  float,
  captionpos=b,
  belowskip=-0.5cm
}

\usepackage{algorithm}
\usepackage{algpseudocode}
\usepackage{float}
\newfloat{algorithm}{t}{top}

\newacronym{simd}{SIMD}{Single Instruction, Multiple Data}
\newacronym{elu}{ELU}{Exponential Linear Unit}
\newacronym{relu}{ReLU}{Rectified Linear Unit}
\newacronym{rpr}{RPR}{Random Partition Relaxation}
\newacronym{mac}{MAC}{Multiply Accumulate}
\newacronym{dma}{DMA}{Direct Memory Access}
\newacronym{cnn}{CNN}{Convolutional Neural Network}
\newacronym{bmi}{BMI}{Brain--Machine Interface}
\newacronym{bci}{BCI}{Brain--Computer Interface}
\newacronym{smr}{SMR}{Sensory Motor Rythms}
\newacronym{eeg}{EEG}{Electroencephalography}
\newacronym{svm}{SVM}{Support Vector Machine}
\newacronym{svd}{SVD}{Singular Value Decomposition}
\newacronym{evd}{EVD}{Eigendecomposition}
\newacronym{iir}{IIR}{Infinite Impulse Response}
\newacronym{fir}{FIR}{Finite Impulse Response}
\newacronym{fc}{FC}{Fabric Controller}
\newacronym{nn}{NN}{Neural Network}
\newacronym{mrc}{MRC}{Multiscale Riemannian Classifier}
\newacronym{flop}{FLOP}{Floating Point Operation}
\newacronym{sos}{SOS}{Second-Order Section}
\newacronym{ipc}{IPC}{Instructions per Cycle}
\newacronym{tcdm}{TCDM}{Tightly Coupled Data Memory}
\newacronym{fpu}{FPU}{Floating Point Unit}
\newacronym{fma}{FMA}{Fused Multiply Add}
\newacronym{alu}{ALU}{Arithmetic Logic Unit}
\newacronym{dsp}{DSP}{Digital Signal Processing}
\newacronym{gpu}{GPU}{Graphics Processing Unit}
\newacronym{soc}{SoC}{System-on-Chip}
\newacronym{mi}{MI}{Motor-Imagery}
\newacronym{csp}{CSP}{Commmon Spatial Patterns}
\newacronym{fbcsp}{FBCSP}{Filter-Bank \acrlong{csp}}
\newacronym{pulp}{PULP}{parallel ultra-low power}
\newacronym{soa}{SoA}{state-of-the-art}
\newacronym{bn}{BN}{Batch Normalization}
\newacronym{isa}{ISA}{Instruction Set Architecture}

\renewcommand{\baselinestretch}{0.92}

\newcommand{\rebuttal}[1]{{\color{black}#1}}

\usepackage{fancyhdr}
\fancypagestyle{mahmood}{%
  \fancyhf{} % clear all fields
  
  \fancyhead[C]{\footnotesize \textcopyright 2020 IEEE. Personal use of this material is permitted.  Permission from IEEE must be obtained for all other uses, in any current or future media, including reprinting/republishing this material for advertising or promotional purposes, creating new collective works, for resale or redistribution to servers or lists, or reuse of any copyrighted component of this work in other works.}
}%
\makeatletter
\let\ps@IEEEtitlepagestyle\ps@mahmood
\makeatother

\begin{document}

\title{Q-EEGNet: an Energy-Efficient 8-bit Quantized Parallel EEGNet Implementation for Edge Motor-Imagery Brain--Machine Interfaces
}

% \author{
% \IEEEauthorblockN{Authors removed for review.}
% \IEEEauthorblockA{}
% }

\author{\IEEEauthorblockN{
    Tibor Schneider\IEEEauthorrefmark{1}, 
    Xiaying Wang\IEEEauthorrefmark{1}, 
    Michael Hersche\IEEEauthorrefmark{1}, 
    Lukas Cavigelli\IEEEauthorrefmark{2},
    Luca Benini\IEEEauthorrefmark{1}\IEEEauthorrefmark{3}}
    \IEEEauthorblockA{\\[-2mm]\IEEEauthorrefmark{1}ETH Zürich, Dept. EE \& IT,  Switzerland \hspace{15mm}\IEEEauthorrefmark{3}University of Bologna, DEI, Italy \\
    \IEEEauthorrefmark{2}Huawei Technologies, Zurich Research Center, Switzerland\vspace{-0.3cm}}
    \thanks{Corresponding emails: \{xiaywang, herschmi\}@iis.ee.ethz.ch}
    }

\maketitle

\begin{abstract}
Motor-Imagery Brain--Machine Interfaces (MI-BMIs) promise direct and accessible communication between human brains and machines by analyzing brain activities recorded with \acrfull{eeg}. 
Latency, reliability, and privacy constraints make it unsuitable to offload the computation to the cloud. 
Practical use cases demand a wearable, battery-operated device with low average power consumption for long-term use. 
Recently, sophisticated algorithms, in particular deep learning models, have emerged for classifying EEG signals. 
While reaching outstanding accuracy, these models often exceed the limitations of edge devices due to their memory and computational requirements.
In this paper, we demonstrate algorithmic and implementation optimizations for \eegnet{}, a compact \acrfull{cnn} suitable for many \acrshort{bmi} paradigms. 
We quantize weights and activations to 8-bit fixed-point with a negligible accuracy loss of 0.4\% on 4-class MI, and present an energy-efficient hardware-aware implementation on the \wolf{} \acrfull{pulp} \acrfull{soc} by utilizing its custom \mbox{RISC-V} ISA extensions and 8-core compute cluster.
With our proposed optimization steps, we can obtain an overall speedup of 64$\times$ and a reduction of up to 85\% in memory footprint with respect to a single-core layer-wise baseline implementation.
Our implementation takes only 5.82\,ms and consumes 0.627\,mJ per inference. 
With 21.0\,GMAC/s/W, it is 256$\times$ more energy-efficient than an \eegnet{} implementation on an ARM Cortex-M7 (0.082\,GMAC/s/W).
%~\cite{Wang2020AnComputing}
\end{abstract}


\section{Introduction}\label{ch:introduction}

A \gls{bmi} is a system that enables direct communication between humans and devices based on signals recorded from brain activities.
One promising \gls{bmi} approach is based on \gls{mi}, which describes the cognitive process of thinking about motions without actually performing them.
Patients with severe physical disabilities could rely on \gls{mi}-\glspl{bmi} to regain independence~\cite{Frolov2017Post-strokeTrial,Kobayashi2018BCI-basedEEG-short}.

%A cheap and widely used method for measuring brain activities is \gls{eeg}. 
MI-BMIs are often based on \gls{eeg}, an accessible and widely used method for measuring brain activities. 
However, \gls{eeg} data show high variability across different subjects as well as different recordings of the same subject, making accurate classification a challenging task. 
%
%Due to the lack of sufficient amount of labeled training data, 
A common approach is to rely on domain-specific knowledge, extracting human-interpretable features such as \gls{fbcsp}~\cite{ang2008filter} or Riemannian geometry features~\cite{Hersche2018_short}. 
Promising alternatives are \glspl{cnn}, which are gaining increasing attention in the MI-BMI field thanks to high \gls{soa} classification accuracy~\cite{schirrmeister2017deep,Lotte2018AUpdate}.
A popular competitor is \eegnet{}~\cite{lawhern2018eegnet}, a CNN-based approach generally applicable to many different \gls{bmi} paradigms, achieving comparable accuracy to architectures tailored to the specific use case while still being compact ($<$3000 parameters), compared to other \glspl{cnn} for MI-BMIs~\cite{li2019densely}. 

Most existing \gls{bmi} systems rely on offline remote computing for classification; however, having those networks mapped to low-cost, low-power embedded platforms, e.g., Microcontroller units (MCUs), is very beneficial.
MCU-based platforms and devices are comfortable, light, and less power-hungry.
Classifying signals on the edge eliminates the latency due to communication, and the energy required for the data transfer.
Besides, processing brain signals on the recording device itself allows users to maintain their privacy.
Several energy-efficient platforms have been proposed in both industry and academia for enabling continuous long-term classification on battery-operated edge devices. 
The most popular energy-efficient MCUs are from the ARM Cortex-M family, with Cortex-M7 being the highest-performing member. 
%
%More recently, academic researchers have proposed an open-source project based on RISC-V \gls{isa}, namely the \gls{pulp} platform. 
Recently, researchers have developed the \gls{pulp} platform based on the RISC-V \gls{isa}~\rebuttal{
% \cite{schiavone2017slow_short, gautschi2017riscy}
\cite{schiavone2017slow_short, gautschi2017riscy}
}, which is built around the concept of using simple cores for energy efficiency, while recovering and scaling up performance through parallelism. 
\gls{pulp} MCUs have proven to outperform the Cortex-M family by at least one order of magnitude in energy efficiency~\cite{garofalo2020pulp, wang2019fann}. In particular, \wolf{}, with its 8-core compute cluster and custom \gls{isa} extensions, can reach up to 274\,GOp/s/W~\cite{pullini2019wolf}.
%Additionally, it is very questionable, regarding privacy, to send personal brain signals to a remote location for further processing.

Nevertheless, both Cortex-M and RISC-V based MCU platforms are tightly constrained both in memory and compute resources, which forced other embedded solutions to tailor and scale down \eegnet{} for the target system resulting in lower classification accuracy~\cite{Wang2020AnComputing}. 
%posing a serious challenge to the embedding of MI-BCI classification on these systems. 
To address this challenge, we present Q-\eegnet{}, an adapted and quantized \eegnet{}~\cite{lawhern2018eegnet} with algorithmic and implementation optimizations to execute \gls{bmi} inference on resource-limited edge devices. 
The proposed methods overcome the necessity of network reduction for embedded implementations and are generally applicable to other \glspl{cnn} in MI-BMIs.
The main contributions of this paper are as follows:
\begin{itemize}
    \item We quantize weights and activations of \eegnet{} from 32-bit float to 8-bit fixed-point representation using quantization-aware training and Random Partition Relaxation (RPR)~\cite{Cavigelli2020RPR:Networks}, resulting in a negligible loss of 0.4\% accuracy on the 4-class BCI Competition IV-2a dataset~\cite{Brunner2008BCIA} (Section~III). This allows the use of vectorized integer operations and the compression of the weights and feature maps by 4$\times$. 
    %Besides, it enables efficient inference on low-power devices without floating-point units.
    \item We present an optimized hardware-aware implementation of the quantized model on \wolf{} (Section~IV). 
    The concurrent execution and the use of the \mbox{RISC-V} \gls{isa} extensions yield a speedup of 36.1$\times$ compared to the baseline single-core implementation. %by overcoming the traditional layer-by-layer computation paradigm, our proposed implementation achieves up to 85\% reduction in memory footprint and an overall speedup of 64$\times$.
    \item \rebuttal{We overcome the traditional layer-by-layer computation paradigm and propose an interleaved implementation that achieves up to 85\% reduction in memory footprint and an overall speedup of 64$\times$.}
    \item Experimental measurements, in Section V, show that the execution of Q-\eegnet{} on \wolf{} takes 5.82\,ms per inference consuming only 0.627\,mJ, yielding an energy-efficiency of 20.957\,GMAC/s/W. 
    Compared to another implementation of a reduced \eegnet{} on an ARM Cortex-M7~\cite{Wang2020AnComputing} with 0.082~GMAC/s/W, Q-\eegnet{} on \wolf{} is 256$\times$ more energy-efficient.  
    %\item With the experimental results, we provide an useful insight on the trade-off between classification accuracy and energy-efficiency for \gls{mi}-\gls{bmi} at the edge.
\end{itemize}
Finally, we release open-source code developed in this work\footnote{https://github.com/pulp-platform/q-eegnet}.

%%% Local Variables:
%%% mode: latex
%%% TeX-master: "report"
%%% End:
\section{Background}\label{ch:background}

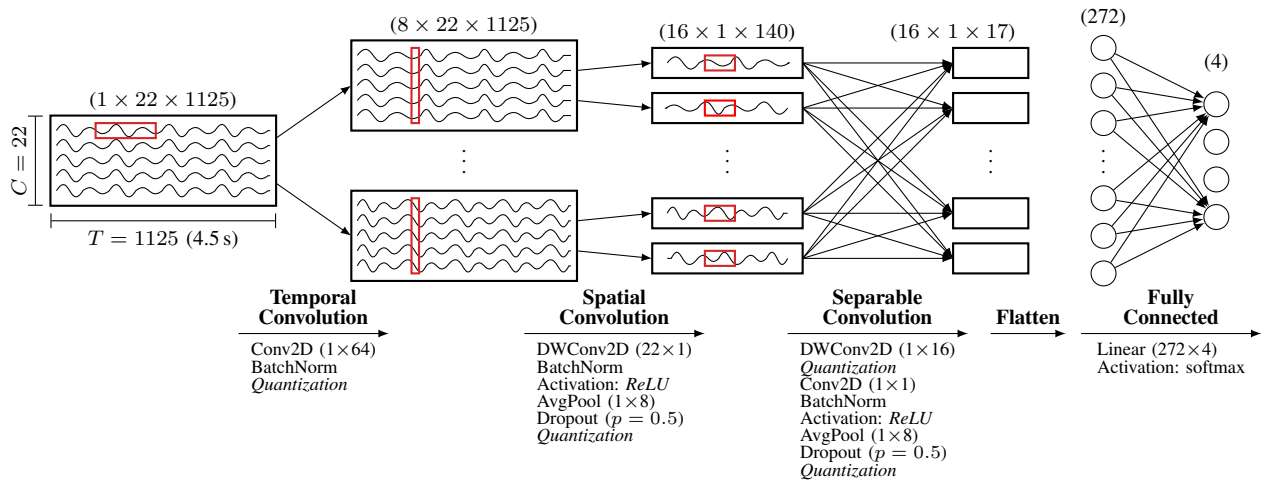
\begin{figure*}[htbp]
  \centering
  \begin{tikzpicture}[scale=1, transform shape]
    %\draw[help lines] (0, 0) grid (16, 8);

    %Layers
    % Temporal Convolution
    \draw[->] (2.5, 1.7) -- (4.5, 1.7) node[pos=0.5, above, align=center, font={\footnotesize \bfseries}] 
      {Temporal\\[-1.5pt]Convolution};
    \node[below, align=left, font=\scriptsize] at (3.5, 1.7) {
      Conv2D (1$\times$64)\\[-1.5pt]
      BatchNorm\\[-1.5pt]
      \textit{Quantization}
    };
    % Spatial Convolution
    \draw[->] (6.3, 1.7) -- (8.7, 1.7) node[pos=0.5, above, align=center, font={\footnotesize \bfseries}] 
      {Spatial\\[-1.5pt]Convolution};
    \node[below, align=left, font=\scriptsize] at (7.5, 1.7) {
      DWConv2D (22$\times$1)\\[-1.5pt]
      BatchNorm\\[-1.5pt]
      Activation: \textit{\acrshort{relu}}\\[-1.5pt]
      AvgPool (1$\times$8)\\[-1.5pt]
      Dropout ($p=0.5$)\\[-1.5pt]
      \textit{Quantization}
    };
    % Separable Convolution
    \draw[->] (9.8, 1.7) -- (12.2, 1.7) node[pos=0.5, above, align=center, font={\footnotesize \bfseries}] 
      {Separable\\[-1.5pt]Convolution};
    \node[below, align=left, font=\scriptsize] at (11, 1.7) {
      DWConv2D (1$\times$16)\\[-1.5pt]
      \textit{Quantization}\\[-1.5pt]
      Conv2D (1$\times$1)\\[-1.5pt]
      BatchNorm\\[-1.5pt]
      Activation: \textit{\acrshort{relu}}\\[-1.5pt]
      AvgPool (1$\times$8)\\[-1.5pt]
      Dropout ($p=0.5$)\\[-1.5pt]
      \textit{Quantization}
    };
    % Flatten
    \draw[->] (12.5, 1.7) -- (13.5, 1.7) node[pos=0.5, above, align=center, font={\footnotesize \bfseries}] 
      {Flatten};
    % Fully Connected
    \draw[->] (13.7, 1.7) -- (16.1, 1.7) node[pos=0.5, above, align=center, font={\footnotesize \bfseries}] 
      {Fully\\[-1.5pt]Connected};
   \node[below, align=left, font=\scriptsize] at (14.9, 1.7) {
  Linear (272$\times$4)\\[-1.5pt]
  Activation: softmax
    };

    % Feature Maps
    %Input
    \draw (1.5, 4.8) node {\footnotesize($1 \times 22 \times 1125$)};
    % L1
    \draw (5.5, 5.8) node {\footnotesize($8 \times 22 \times 1125$)};
    % L2
    \draw (9, 5.7) node {\footnotesize($16 \times 1 \times 140$)};
    % L3
    \draw (12, 5.7) node {\footnotesize($16 \times 1 \times 17$)};
    % Flat
    \draw (14, 5.9) node {\footnotesize(272)};
    % output
    \draw (15.5, 5.3) node {\footnotesize(4)};

    % input featuremap
    \begin{scope}[shift={(0, 3)}]
      \draw[thick](0, 0.4) rectangle (3, 1.6);
      \foreach \y in {0.6, 0.8, 1.0, 1.2, 1.4}
        \draw[snake=eeg, segment length=20pt] (0.08, \y) -- (2.92, \y);
      \draw[thick, color=matplotlib1] (0.6, 1.3) rectangle (1.4, 1.5);

      \draw[|-|] (-0.2, 0.4) -- (-0.2, 1.6) node[pos=0.5, above, rotate=90]{\footnotesize$C=22$};
      \draw[|-|] (0, 0.2) -- (3, 0.2) node[pos=0.5, below]{\footnotesize$T=1125$ (4.5\,s)};
    \end{scope}

    \draw[->](3, 4.3) -- (4, 5);
    \draw[->](3, 3.7) -- (4, 3);

    % L1 featuremap
    \begin{scope}[shift={(4, 4)}]
      \draw[thick](0, 0.4) rectangle (3, 1.6);
      \foreach \y in {0.6, 0.8, 1.0, 1.2, 1.4}
        \draw[snake=eeg filter1, segment length=20pt] (0.08, \y) -- (2.92, \y);
      \draw[thick, color=matplotlib1] (0.8, 0.5) rectangle (0.9, 1.5);
      \draw[->] (3, 0.8) -- (4, 0.7);
      \draw[->] (3, 1.2) -- (4, 1.3);
    \end{scope}
    \draw (5.5, 4.1) node {\footnotesize $\vdots$};
    \begin{scope}[shift={(4, 2)}]
      \draw[thick](0, 0.4) rectangle (3, 1.6);
      \foreach \y in {0.6, 0.8, 1.0, 1.2, 1.4}
        \draw[snake=eeg filter2, segment length=20pt] (0.08, \y) -- (2.92, \y);
      \draw[thick, color=matplotlib1] (0.8, 0.5) rectangle (0.9, 1.5);
      \draw[->] (3, 0.8) -- (4, 0.7);
      \draw[->] (3, 1.2) -- (4, 1.3);
    \end{scope}

    % L2 Featuremap
    \begin{scope}[shift={(8, 5.3)}]
      \draw[thick](0, -0.2) rectangle (2, 0.2);
      \draw[snake=eeg filter1, segment length=20pt] (0.2, 0) -- (1.8, 0);
      \draw[thick, color=matplotlib1] (0.7, -0.1) rectangle (1.1, 0.1);
    \end{scope}
    \begin{scope}[shift={(8, 4.7)}]
      \draw[thick](0, -0.2) rectangle (2, 0.2);
      \draw[snake=eeg filter1, segment length=20pt] (1.8, 0) -- (0.2, 0);
      \draw[thick, red] (0.7, -0.1) rectangle (1.1, 0.1);
    \end{scope}
    \draw (9, 4.1) node {\footnotesize $\vdots$};
    \begin{scope}[shift={(8, 3.3)}]
      \draw[thick](0, -0.2) rectangle (2, 0.2);
      \draw[snake=eeg filter2, segment length=20pt] (0.2, 0) -- (1.8, 0);
      \draw[thick, color=matplotlib1] (0.7, -0.1) rectangle (1.1, 0.1);
    \end{scope}
    \begin{scope}[shift={(8, 2.7)}]
      \draw[thick](0, -0.2) rectangle (2, 0.2);
      \draw[snake=eeg filter2, segment length=20pt] (1.8, 0) -- (0.2, 0);
      \draw[thick, color=matplotlib1] (0.7, -0.1) rectangle (1.1, 0.1);
    \end{scope}

    % L3 Featuremap
    \begin{scope}[shift={(12, 5.3)}]
      \draw[thick](0, -0.2) rectangle (1, 0.2);
    \end{scope}
    \begin{scope}[shift={(12, 4.7)}]
      \draw[thick](0, -0.2) rectangle (1, 0.2);
    \end{scope}
    \draw (12.5, 4.1) node {\footnotesize $\vdots$};
    \begin{scope}[shift={(12, 3.3)}]
      \draw[thick](0, -0.2) rectangle (1, 0.2);
    \end{scope}
    \begin{scope}[shift={(12, 2.7)}]
      \draw[thick](0, -0.2) rectangle (1, 0.2);
    \end{scope}
    
    % separable convolution arrows
    \foreach \a in {2.7, 3.3, 4.7, 5.3}
      \foreach \b in {2.7, 3.3, 4.7, 5.3}
        \draw[->] (10, \a) -> (12, \b);

    % flatten
    \begin{scope}[shift={(14, 6)}]
      \node[circle, draw] at (0, -0.5) (fl1) {};
      \node[circle, draw] at (0, -1.0) (fl2) {};
      \node[circle, draw] at (0, -1.5) (fl3) {};
      \node at (0, -1.9) {\footnotesize $\vdots$};
      \node[circle, draw] at (0, -2.5) (fl4) {};
      \node[circle, draw] at (0, -3.0) (fl5) {};
      \node[circle, draw] at (0, -3.5) (fl6) {};
    \end{scope}

    % FC
    \begin{scope}[shift={(15.5, 4.75)}]
      \node[circle, draw] at (0, -0.0) (out0) {};
      \node[circle, draw] at (0, -0.5) (out1) {};
      \node[circle, draw] at (0, -1.0) (out2) {};
      \node[circle, draw] at (0, -1.5) (out3) {};
    \end{scope}

    % FC connections
    \foreach \s in {fl1, fl2, fl3, fl4, fl5, fl6} {
      % \foreach \e in {out0, out1, out2, out3} {
      \foreach \e in {out0, out3} {
        \draw[->] (\s) -- (\e);
      }
    }

  \end{tikzpicture}
  \vspace{-0.3cm}
  \caption{Q-\eegnet{} architecture applied to BCI Competition IV-2a dataset~\cite{lawhern2018eegnet}. The input signal is quantized to 8-bit fixed-point.}\label{fig:background:eegnet}
  \vspace{-0.3cm}
\end{figure*}

\subsection{Dataset description}\label{ch:background:dataset}
In this work, we use the BCI Competition IV-2a dataset~\cite{Brunner2008BCIA}, which contains recordings from 9 different subjects and distinguishes between four classes of imagined movements: left and right hand, both feet, and the tongue.
22 different \gls{eeg} channels were recorded, sampled at 250\,Hz.
The data is pre-processed with a bandpass filter between 0.5 and 100\,Hz.
Each subject completed two recording sessions on two different days.
Recordings from the first day are used only for training, and samples from the second session are used exclusively for testing.
Per subject and per session, 288 trials were recorded, of which almost 10\% were excluded due to artifacts originating mostly from eye movements.
The dataset, however, remains balanced.
Per trial, 6\,s of \gls{eeg} data is recorded: 2\,s before the MI-cue, 1\,s of showing the cue, and 3\,s when the subject was executing MI.

\subsection{EEGNet}\label{ch:background:eegnet}

\eegnet{}~\cite{lawhern2018eegnet} is a \acrfull{cnn} designed to apply to many different \gls{bmi} paradigms such as P300 event-related potential (P300), feedback error-related negativity (ERN), movement-related cortical potential (MRCP), and sensory-motor rhythm (SMR) encountered in MI.  
Another design goal of \eegnet{} is to contain as few model parameters as possible, which is essential in many applications due to the limited amount of labeled training data.
%
%The network applied on the 4-class MI EEG data of the BCI Competition IV-2a is depicted in Fig.~\ref{fig:background:eegnet}.
%
It consists of three convolutional layers in the Temporal, Spatial, and Separable Convolution blocks, depicted in Fig.~\ref{fig:background:eegnet}.
Each convolution is followed by a \gls{bn} layer and a linear or \gls{elu} activation. 
All convolutional kernels are 1-dimensional (1D).
The network contains two average pooling layers to reduce the size of the feature maps.
The final classification is a linear fully-connected (FC) layer.
Thanks to the use of depth-wise convolutions and pooling layers, \eegnet{} requires only 2548 parameters and 13.14 million\footnote{Each convolutional layer contributes with $h_{out}\cdot w_{out}\cdot n_{out}  \cdot h_k \cdot w_k \cdot n_{in}/g$ MACs, with $g$ being the number of groups (commonly $g$=1, for depth-wise sep. conv. $g$=$n_{in}$)} \gls{mac} operations per inference.
Nevertheless, it achieves an accuracy of 71.0\% on 4-class MI, which is 3\% more accurate than the winner of the BCI competition IV-2a~\cite{ang2008filter}. 
% Changed from 2544 to 2548 parameters (adding bias at FC)
% Changed number of MACs to 13.14

\subsection{\wolf{}}\label{ch:background:wolf}

\wolf{}~\cite{pullini2019wolf} is a \gls{soc} for embedded, low-power applications.
\wolf{} is split into two computation domains: the \gls{soc} domain and the compute cluster.
The \gls{soc} domain is responsible for handling inputs and outputs, as well as computationally simple tasks.
It is based around the fabric controller
%, a 12\,kgates, two-pileline stage 
with a \riscv{} processor called \ibex{}~\cite{schiavone2017slow_short}.
The \gls{soc} domain contains 448\,kB of shared L2 memory. % distributed on four memory banks.
The compute cluster consists of eight 
in-order four-stage \riscv{} RV32IMFC\xpulp{} processors called 
\riscy{}~\cite{gautschi2017riscy} (now maintained by the OpenHW Group as CV32E40P), which support the RVC32IMF instruction set and the \xpulp{} extension, adding support for \gls{simd}, load and store post-increment, and hardware loops.
The cluster is available on demand; individual cores can be disabled to save energy.
All cores have access to 64\,kB of shared L1 memory via the \gls{tcdm} interconnect.
%The L1 memory is distributed among 16 memory banks.
%The \gls{tcdm} interconnect is able to serve each of the eight cores in the same cycle, assuming that all cores access different memory banks.
%A memory access takes two cycles (assuming no \gls{tcdm} contentions), but the latency is hidden when the data is not immediately used in the next cycle.
A \gls{dma} controller is responsible for moving data between L1 and L2 memory.

\rebuttal{\section{Related work}} \label{ch:background:related-work}
As a result of the emerging Internet of Things (IoT), which brings intelligence close to the sensor, the current literature is rich in implementing inference of neural networks on low-power edge devices and is also gaining increasing attention in MI-BCI.
\textsc{Cube.AI} converts trained models from Keras and generates an optimized code
%, including quantization, 
for several embedded platforms of the STM32 series.
In contrast, \textsc{TensorFlow Lite} supports various platforms, including \riscv{}~\cite{louis2019towards}.
However, the resulting implementation for \riscv{} does not support parallel execution.
\textsc{FANN-on-MCU}~\cite{wang2019fann} is a different framework for exporting optimized neural networks to ARM processors, and to \pulp{}-based systems.
However, this framework does not offer convolutional layers required for \eegnet{}.
\textsc{PULP-NN}~\cite{garofalo2020pulp} is a library containing highly optimized implementations for typical (convolutional) neural networks targeting the \pulp{}-platform.

In~\cite{Wang2020AnComputing}, \eegnet{} was applied to the Physionet Motor Movement/Imagery Dataset,
%~\cite{goldberger2000physiobank}
achieving \gls{soa} accuracy.
The model was quantized and ported to an ARM Cortex-M7 using \textsc{Cube.AI}, i.e., the X-CUBE-AI expansion package of STM32CubeMX. 
However, the current package expansion can quantize only the FC layer to 8 bits, which is almost insignificant in terms of computation compared to the rest of \eegnet{}.
The input feature map had to be scaled down significantly from (64 channels $\times$ 480 time-samples) to (38 channels $\times$ 80 time-samples)  by sub-sampling, EEG channel reduction, and narrowing the time window, such that the feature maps fitted on the available SRAM.

%%% Local Variables:
%%% mode: latex
%%% TeX-master: "report"
%%% End:

\section{Model Design and Quantization}
This section explains how \eegnet{} is modified, quantized, and trained, resulting in Q-\eegnet{} targeting a low-power implementation.
Fig.~\ref{fig:background:eegnet} illustrates the layers of Q-\eegnet{}, which processes 4.5\,s of EEG data, starting 0.5\,s before the onset of the MI-cue according to the timing scheme of the BCI Competition IV-2a dataset. 
%
%To improve energy efficiency without affecting accuracy, w
We have modified the original \eegnet{} as follows:

\begin{itemize}
    % \item The 1D pooling in time dimension is increased from 4 to 8 in both the spatial and the separable convolution blocks to reduce the size of feature maps in the network. % without affecting accuracy.
    \item The computationally expensive \gls{elu} activations are replaced with \gls{relu}.
    \item The weight regularization is removed from the training procedure since it has no effect on the accuracy and interferes with the quantization procedure. 
\end{itemize}

\rebuttal{
%Works in literature [\note{cite}] have shown that it is possible to quantize neural networks to 8- or lower bit representation without significant loss in accuracy.}
In this work, all weights and activations, including the input signals, are quantized independently to 8-bit fixed-point representations. % (i.e., the dynamic range $S_\xi$ is chosen only based on the respective matrix).
This reduces the memory footprint and enables maximal use of the underlying microprocessor architecture with its 4-way SIMD instructions.}

As shown in Fig.~\ref{fig:background:eegnet}, we do not introduce quantization between every single layer of the network.
Instead, we requantize only before the convolutional and the FC layers.
The reason is that all other layers (i.e., \gls{bn}, \gls{relu}, and average pooling layers) are defined locally.
They can easily be computed one after the other, without writing back to memory.
Requantizing those values to less than 32-bit fixed-point values would increase the quantization error and introduce a higher overhead than the subsequent speedup.

% Thus, treating those activations as 32-bit fixed-point values does not impact the energy-efficiency.
% Doing more requantization would introduce more quantization error, resulting in a model that is more difficult to train.

A quantization layer first rescales the activations according to their expected range, and then reduces the precision from 32-bit to 8-bit fixed-point. 
Usually, it is beneficial to choose the scaling factor to be a power of two, such that it can be implemented with an efficient bit-shift instead of an expensive integer division.
However, the scaling factors and offsets of the \gls{bn} layers are learned during training, and cannot be approximated as powers of two. Thus, we require a full integer division.
Alternatively, the \gls{bn} layer could be merged into the preceding convolution before the quantization-aware retraining and the shift be constrained to a power-of-two. However, this removes a degree of freedom and might thus adversely affect the final accuracy.
%However, we use a training method which is aware of the quantization.
%
%Hence, embedding the \gls{bn} transformation into the weights after training removes the benefits of such sophisticated training methods.

\begin{figure}
  \centering
  \resizebox{\columnwidth}{!}{%
  \begin{tikzpicture}
    \begin{axis}[
      no markers,
      width=0.9\columnwidth, height=0.65\columnwidth,
      xlabel = {\scriptsize Epoch},
      ylabel = {\scriptsize Loss},
      ylabel near ticks,
      xtick = {0, 150, 300, 450, 550, 650},
      ticklabel style={font={\scriptsize}},
      xmin = 0, xmax = 650,
      ymin = 0, ymax = 1.5,
      axis y line* = right]

      \addplot+ [thick] table [x=epoch, y=train_loss, col sep=comma] {measurements/training_S5.csv};
      % \addplot+ [thick] table [x=epoch, y=valid_loss, col sep=comma] {measurements/training_S5.csv};

    \end{axis}
    \begin{axis}[
      no markers,
      width=0.9\columnwidth, height=0.65\columnwidth,
      ylabel = {\scriptsize Accuracy / \gls{rpr} fraction},
      ylabel near ticks,
      xmin = 0, xmax = 650,
      ymin = 0, ymax = 1,
      %hide x axis,
      axis y line* = left,
      xtick = {0, 150, 300, 450, 600},
      xticklabels = {,,,,},
      grid = both,
      ticklabel style={font={\scriptsize}},
      legend style={
        at={(0.5, 1.05)},
        anchor=south,
        draw=none,
        /tikz/column 2/.style={column sep=10pt},
        /tikz/column 4/.style={column sep=10pt},
        font={\scriptsize}
      },
      legend columns = 2]

      \addplot+ [thick] coordinates {(0, 0)};

      \addplot+ [thick, const plot, restrict expr to domain={\coordindex}{548:650}] table [x=epoch, y=INQ/fraction, col sep=comma] {measurements/training_S5.csv};

      \addplot+ [thick] table [x=epoch, y=train_metric, col sep=comma] {measurements/training_S5.csv};

      \addplot+ [thick] coordinates {(450, 0) (450, 1)};

      \legend{Training Loss, \gls{rpr} Fraction, Training Accuracy, Activation Quantization};
    
    \end{axis}
  \end{tikzpicture}
  }
  \vspace{-.8cm}
  \caption{Training loss and accuracy of Q-\eegnet{} on Subject~5.}% of the BCI Competition IV-2a dataset.}
  \vspace{-.3cm}
  \label{fig:eegnet:design:training}
\end{figure}
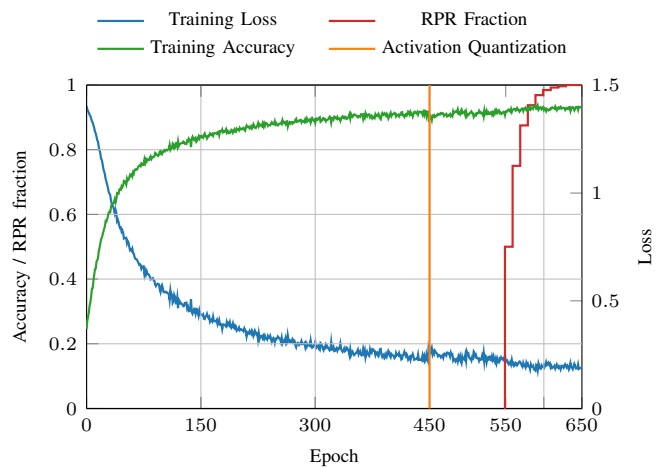

The network is first trained in full precision, for 450 epochs, to get a pre-trained model.
%(The original \eegnet{} is trained for 500 epochs.)
In the 450th epoch, the value range of the activations is monitored. 
In subsequent epochs, they are quantized using the straight-through estimator (STE), i.e., the values are quantized in the forward pass while the full precision values are used for backpropagation~\cite{jacob2018quantization}.
The next 100 epochs are necessary for the network to adapt to the quantized activations.
During the last 100 epochs of the training process, the weights of the network are quantized incrementally using \gls{rpr}~\cite{Cavigelli2020RPR:Networks}.
Fig.~\ref{fig:eegnet:design:training} illustrates the training process and shows the training loss and accuracy for Subject 5.

% added back this figure

%\subsection{Implementation}\label{ch:eegnet:implementation}
\section{Implementation and Optimizations}\label{ch:eegnet:implementation}

This section elaborates on implementing \mbox{Q-\eegnet{}} on \wolf{} and highlights our novel optimizations, which enable high energy-efficiency.

\eegnet{} contains multiple 2D convolutions.
However, the kernel size in all layers spans only one dimension, which means that they can be computed using exclusively 1D convolutions.
We start with a baseline implementation on a single \riscy{} core, not utilizing the \gls{simd} instructions.
For the baseline, the weights and feature maps are transferred layer-by-layer from L2 to L1 memory using the \gls{dma} unit; the computation is done after every completed transfer, and the result is subsequently moved back to L2 memory. 
On top of this, we incrementally add the following optimization steps:

\subsection{SIMD and loop unrolling}\label{ch:implementation:simd}
Since all activations and weights are quantized to 8 bits, four of them are packed into a single 32-bit word to have much more effective loads and exploit \gls{simd} instructions~\rebuttal{\cite{gautschi2017riscy}}. 
Moreover, loop unrolling is applied to reduce the pipeline stalls after load operations.  %to hide the latency in the memory access.
For 1D convolutions, we use the optimized implementation from the PULP-DSP library~\cite{wang2019dsp}, which computes four elements of the output vector at a time using 8-bit operands and 32-bit accumulators.
%It is now to such a word with four values at the same time perform computations using...

Based on \wolf{}'s architecture and the \gls{simd} instructions available on the \riscy{} cores, we choose the time dimension to be the innermost dimension to exploit the optimized 1D convolutions. %, along which the data is packed.
%This is different to traditional implementations of \glspl{cnn} on \glspl{gpu}, where the inner most dimension is chosen to be the different channels.
Additionally, we align every feature map and weight matrix to 4 Bytes, eliminating all misaligned memory accesses.

\subsection{Transpose feature maps}\label{ch:implementation:transpose}

% The kernel of the spatial convolution only spans the space dimension.
The kernels of the Spatial Convolution span only the space dimension (along the different \gls{eeg} channels), whereas the feature maps are packed along the time dimension.
This prohibits the use of \gls{simd} for this layer.
However, transposing the feature maps (switching space and time dimension) allows \gls{simd} instructions for the convolution, analogous to the Temporal Convolution.
Additionally, since the kernel size is equal to the number of \gls{eeg} channels, the Spatial Convolution can be implemented as a series of dot products.

Similarly, the pointwise convolution (the second part of the Separable Convolution) only consists of $(1\times1)$ kernels.
By switching the time dimension with the channel dimension, we can compute the convolution as a series of \gls{simd} dot products.

\subsection{Concurrent execution}\label{ch:implementation:concurrent}
The compute cluster of the target platform contains eight cores; using all of them for inference has a significant impact on the performance and the energy efficiency.
%All eight cores of \wolf{} are fully utilized by distributing all $22 \times 8 = 176$ 
%different 1D convolutions among all cores in the cluster 
For the Temporal Convolution, all the $22 \times 8 = 176$ different 1D convolutions are distributed among the eight cores.
%For the spatial convolution, concurrent execution is achieved by assigning each core of the cluster a different input channel.
The Spatial Convolution is computed in parallel by assigning each core of the cluster one of the eight feature maps.
%The depthwise convolution with a 1D kernel is operationally identical to a dot product, where the weights can be reused.
The analogous is done for both convolutional layers in the Separable Convolution.
%For the separable convolution, ...
Finally, we implement the FC layer on a single core, since it contains less than 0.01\% of all \gls{mac} operations in the entire network; therefore, any improvement at this stage has no significant effect on the overall performance.
%Finally, the linear layer is computed linearly, since its computation takes less than 0.3\% of the total compute time of the network.

\subsection{Cross-correlation instead of convolution}\label{ch:eegnet:implementation:xcorr}

Cross-correlation is a close sibling to convolution.
The only difference between those operations is that, for the convolution, the weight vector must be flipped.
% \begin{align}
%   y[k] &= \text{Conv}(x, w)[k] = \sum_{i=1}^N x[k + i] \cdot w[N - i] \\
%   y[k] &= \text{Corr}(x, \tilde{w})[k] = \sum_{i=1}^N x[k + i] \cdot \tilde{w}[i] 
% \end{align}
Computing a convolution requires reversing the weight vector, resulting in an additional instruction in the innermost loop.
By storing the weight vector in reverse order on the target, and using cross-correlations instead of the convolutions, this shuffle instruction is no longer necessary, reducing the number of instructions. % in the innermost loop.

\subsection{Interleaved layers}\label{ch:implementation:interleaved} %temporal and spatial convolutions
The Temporal Convolution creates eight different output channels, increasing the size of the feature maps by 8$\times$.
The Spatial Convolution then reduces the number of \gls{eeg} channels from 22 down to 1, and the subsequent pooling layer reduces the time resolution by a factor of 8$\times$.
Therefore, not storing the output of the Temporal Convolution would reduce the total memory requirements of Q-\eegnet{} by 85\%.

This reduction in memory can be achieved by exploiting the nature of convolution layers; a small input region fully determines a single output feature.
In the case of the Spatial Convolution, an output feature can be computed from a single column (spatial dimension) at the input feature map.
%, thanks to the use of depth-wise convolution in \eegnet{}.
Thus, we interleave the computation of the Temporal and Spatial Convolutions, computing only a single column with the Temporal Convolution, followed by computing a single output element of the Spatial Convolution.
For the intermediate result, we reuse the same memory location.

\subsection{Merging batch normalization}\label{ch:implementation:collect_bn}
A trained \gls{bn} layer can be computed in several different ways.
For a fixed-point implementation, the most precise results are achieved by adding a bias $b$ and then dividing by a factor $f$.
The \gls{bn} layer in the Temporal Convolution block of Fig.~\ref{fig:background:eegnet} requires almost 200\,000 integer divisions.
To reduce the number of divisions, which can be very costly on low-power embedded platforms, we exploit the linearity of the convolution operation.
More specifically, the division of the first \gls{bn} is moved after the depth-wise convolution and combined with the \gls{bn} in the Spatial Convolution block, reducing the number of divisions by more than a factor 10$\times$.
This can be expressed as:
% \begin{equation*}
%   y = \frac{\frac{x \ast w_T + b_1}{f_1} \ast w_S+ b_2}{f_2} = \frac{(x \ast w_T + b_1) \ast w_S + f_1 b_2}{f_1 f_2},
% \end{equation*}

\medmuskip=2mu
\thinmuskip=2mu
\thickmuskip=2mu
\begin{equation*}
  y = \frac{(x \star w_T + b_1)/f_1 \star w_S+ b_2}{f_2} = \frac{(x \ast w_T + b_1) \star w_S + f_1 b_2}{f_1 f_2},
\end{equation*}
where $\star$ is the convolution operation, $x$ the input feature map, $w_T$ the network weights in the Temporal Convolution block, and $w_S$ the weights in the Spatial Convolution block.
$b_1$ and $f_1$ represent the bias and the normalization factor of the first \gls{bn} layer, similarly, $b_2$ and $ f_2$ for the second \gls{bn} layer.
%We can see that divisions are now only required after the Spatial convolution, where the number of activations is already more than 10 times lower.

Note that after the Temporal Convolution, the resulting features are in 32-bit representation using the 1D convolution of PULP-DSP library. 
Since the complexity of the depth-wise convolution in the Spatial Convolution block is orders of magnitude lower than the Temporal Convolution, we do not requantize the activations and execute the depth-wise convolution in 32-bit with negligible impact on the overall performance. 
This also reduces the error introduced by the requantization.

\subsection{Layer reordering}\label{ch:implementation:reordering}

In both the Spatial and Separable Convolution blocks, the convolution is followed by a \gls{bn}, a \gls{relu}, and a pooling layer.
However, it is beneficial first to execute the pooling layer to reduce the feature maps, and then apply the other layers, which decreases the overall number of operations.
In contrast to the non-linear \gls{relu} layer, the \gls{bn} can be computed after the pooling layer, shown as follows:
\begin{equation*}
    y = \frac{1}{N} \sum_{i=0}^{N-1} \max\left(\frac{x_i + b}{f}, 0\right) = \frac{Nb + \sum \max(x_i, -b)}{Nf},
\end{equation*}
where $b$ and $f$ are respectively the bias and the normalization factor of \gls{bn}, the $\max( \cdot , 0)$ is the original ReLU activation, and the average summation represents the pooling layer.
The new \gls{relu} activation $\max( \cdot , -b)$ is shifted by $b$, and the \gls{bn} is combined with the division from the average pooling layer.
%For this, we change the \gls{relu} to be $f(x) = \max(x, -B)$, where $B$ is the bias of the \gls{bn} layer.
%Additionally, we can integrate the division (i.e., bit-shift) of the average pooling layer into the division already necessary for the \gls{bn}.
This reduces the number of divisions by the pooling factor $N$ and the number of additions by $N-1$.
%Additionally, the quantization error is reduced, since the pooling is computed in higher precision.

%Finally, to reduce the number of required loads, the \gls{relu}, pooling and \gls{bn} layers can be computed while the result is still in local registers from the previous computation.

\subsection{Replicate feature maps}
%\paragraph{Duplicating the Feature Map}
\label{ch:eegnet:implementation:duplication}
In order to use \gls{simd} for computing convolutions, we need to re-shuffle the data whenever the kernel is shifted by 1 Byte, as can be seen in Listing~\ref{lst:eegnet:conv:shuffle}, because the packed data access is no longer aligned.
However, the additional shuffle instructions can be avoided by replicating feature maps. 
We use the \gls{dma} to copy the feature maps four times to local L1 memory, each of which is shifted by 1 Byte. 
These \gls{dma} transfers add an insignificant overhead, compared to the shuffle instructions they replace.
%The baseline implementation of convolutions using \gls{simd} 
%(see Listing~\ref{lst:eegnet:conv:shuffle}) 
%requires shuffle instructions.
%The reason for this is that the kernel is shifted by a single element (1 Byte instead of 1 Word), which is then no longer aligned to the packed format by which the data is stored in memory.
% The reshuffling, however, does not need to be computed in the innermost loop of the convolution but can be prepared with \gls{dma} transfers.
% The feature map can be copied to local L1 memory four times, always shifted by 1 Byte.
The resulting implementation no longer requires shuffle instructions, as shown in Listing~\ref{lst:eegnet:conv:data_dup}, at the cost of more memory usage and accesses.
%(see Listing~\ref{lst:eegnet:conv:data_dup})
%Table~\ref{tab:eegnet-implementation-dupinp} shows the potential improvement of using this method compared to the baseline approach.
The L1 memory available inside the cluster is not large enough to fit the entire input data, when replicated four times.
Hence, we split the data into five similarly sized parts along the time dimension,  \rebuttal{which allowed us to fit the data into L1 memory and at the same time minimize the number of DMA transfers}. %
%This results in nine different regions, some of which spawn two separate parts.
The DMA independently transfers the data into the cluster memory while the cores are computing on the previously loaded data,  
%Inside a non-overlapping region, the \gls{dma} transfers the data for the next region into the cluster memory, 
reducing the idle time of the cores.

\begin{lstlisting}[
  style=RISCV_STYLE,
  caption={Cross-correlation with shuffle. The pointer to the weights is stored in register \lstinline!a7!, and the pointer to the data in register \lstinline!a8!. Registers \lstinline!a9!, \lstinline!a10! and \lstinline!a11! contain the appropriate shuffle mask.},
  label={lst:eegnet:conv:shuffle}]
lp.setup 0,a2,end
    p.lw t4,4(a7!)          // t4 = {w0,w1,w2,w3}
    p.lw t0,4(a8!)          // t0 = {x0,x1,x2,x3}
    lw t3,a4                // t3 = {x4,x5,x6,x7}
    mv t1,t0
    mv t2,t0
    pv.shuffle2.b t1,t3,a9  // t1 = {x1,x2,x3,x4}
    pv.shuffle2.b t2,t3,a10 // t2 = {x2,x3,x4,x5}
    pv.shuffle2.b t3,t0,a11 // t3 = {x3,x4,x5,x6}
    pv.sdotsp.b t0,t4,a3    // a3 += w0*x0+w1*x1+w2*x2+w3*x3
    pv.sdotsp.b t1,t4,a4    // a4 += w0*x1+w1*x2+w2*x3+w3*x4
    pv.sdotsp.b t2,t4,a5    // a5 += w0*x2+w1*x3+w2*x4+w3*x5
end:pv.sdotsp.b t3,t4,a6    // a6 += w0*x3+w1*x4+w2*x5+w3*x6
\end{lstlisting}

\begin{lstlisting}[
  style=RISCV_STYLE,
  firstnumber=0,
  caption={Cross-correlation with data replication. The pointer to the weights is stored in register \lstinline!a7!, while the pointer to the 4 copies of the data (shifted by 1 Byte) are stored in registers \lstinline!a8!, \lstinline!a9!, \lstinline!a10!, and \lstinline!a11!, respectively.},
  label={lst:eegnet:conv:data_dup}]
lp.setup 0,a2,end
    p.lw t4,4(a7!)         // t4 = {w0,w1,w2,w3}
    p.lw t0,4(a8!)         // t0 = {x0,x1,x2,x3}
    p.lw t1,4(a9!)         // t1 = {x1,x2,x3,x4}
    p.lw t2,4(a10!)        // t2 = {x2,x3,x4,x5}
    p.lw t3,4(a11!)        // t3 = {x3,x4,x5,x6}
    pv.sdotsp.b t0,t4,a3   // a3 += w0*x0+w1*x1+w2*x2+w3*x3
    pv.sdotsp.b t1,t4,a4   // a4 += w0*x1+w1*x2+w2*x3+w3*x4
    pv.sdotsp.b t2,t4,a5   // a5 += w0*x2+w1*x3+w2*x4+w3*x5
end:pv.sdotsp.b t3,t4,a6   // a6 += w0*x3+w1*x4+w2*x5+w3*x6
\end{lstlisting}

%%% Local Variables:
%%% mode: latex
%%% TeX-master: "report"
%%% End:
\section{Experimental Results}\label{ch:results}

\begin{table}
  \centering
  \caption{Classification Accuracy (\% Avg. $\pm$ Std. Dev. Over 50 Runs) on 4-Class BCI Competition IV-2a in Full Precision and Quantized}\label{tab:results:accuracy}
  \vspace{-.2cm}
  {
  \begin{threeparttable}
    \footnotesize
    \begin{tabular}{lccc}
      \toprule
      & \multicolumn{2}{c}{\eegnet{}} & Q-\eegnet{} \\
       \cmidrule(r){1-1}
      \cmidrule(r){2-3}
      \cmidrule(r){4-4}
      Activation & ELU & ReLU & ReLU\\
      Quantization & none & none & 8 bits\\
      \cmidrule(r){1-1}
      \cmidrule(r){2-3}
      \cmidrule(r){4-4}
        % Subject 1 & 80.9 & 80.7 & 81.0 \\
        % Subject 2 & 56.9 & 53.1 & 51.6 \\
        % Subject 3 & 87.1 & 91.6 & 91.4 \\
        % Subject 4 & 60.9 & 58.5 & 58.5 \\
        % Subject 5 & 70.5 & 66.7 & 67.8 \\
        % Subject 6 & 54.2 & 52.3 & 50.4 \\
        % Subject 7 & 74.8 & 75.9 & 75.6 \\
        % Subject 8 & 78.0 & 80.2 & 81.9 \\
        % Subject 9 & 76.0 & 79.4 & 79.3 \\
        %Subject 1 & 80.9 \rebuttal{$\pm$ 2.5} & 80.7 \rebuttal{$\pm$ 2.2} & 81.0 \rebuttal{$\pm$ 3.0} \\
        %Subject 2 & 56.9 \rebuttal{$\pm$ 3.4} & 53.1 \rebuttal{$\pm$ 4.4} & 51.6 \rebuttal{$\pm$ 3.3} \\
        %Subject 3 & 87.1 \rebuttal{$\pm$ 2.6} & 91.6 \rebuttal{$\pm$ 2.1} & 91.4 \rebuttal{$\pm$ 2.1} \\
        %Subject 4 & 60.9 \rebuttal{$\pm$ 3.6} & 58.5 \rebuttal{$\pm$ 4.0} & 58.5 \rebuttal{$\pm$ 4.6} \\
        %Subject 5 & 70.5 \rebuttal{$\pm$ 2.4} & 66.7 \rebuttal{$\pm$ 3.2} & 67.8 \rebuttal{$\pm$ 2.6} \\
        %Subject 6 & 54.2 \rebuttal{$\pm$ 2.4} & 52.3 \rebuttal{$\pm$ 3.8} & 50.4 \rebuttal{$\pm$ 3.4} \\
        %Subject 7 & 74.8 \rebuttal{$\pm$ 5.3} & 75.9 \rebuttal{$\pm$ 5.2} & 75.6 \rebuttal{$\pm$ 4.8} \\
        %Subject 8 & 78.0 \rebuttal{$\pm$ 3.6} & 80.2 \rebuttal{$\pm$ 2.1} & 81.9 \rebuttal{$\pm$ 1.4} \\
        %Subject 9 & 76.0 \rebuttal{$\pm$ 5.1} & 79.4 \rebuttal{$\pm$ 3.1} & 79.3 \rebuttal{$\pm$ 2.7} \\
        Subject 1 & 81.0 \rebuttal{$\pm$ 2.4} & 81.1 \rebuttal{$\pm$ 2.2} & 81.0 \rebuttal{$\pm$ 2.3} \\
        Subject 2 & 57.6 \rebuttal{$\pm$ 4.5} & 52.2 \rebuttal{$\pm$ 4.4} & 53.1 \rebuttal{$\pm$ 4.2} \\
        Subject 3 & 87.9 \rebuttal{$\pm$ 2.7} & 91.3 \rebuttal{$\pm$ 2.1} & 91.2 \rebuttal{$\pm$ 2.4} \\
        Subject 4 & 61.6 \rebuttal{$\pm$ 3.4} & 59.1 \rebuttal{$\pm$ 4.0} & 58.1 \rebuttal{$\pm$ 4.0} \\
        Subject 5 & 70.6 \rebuttal{$\pm$ 2.3} & 68.6 \rebuttal{$\pm$ 3.2} & 68.4 \rebuttal{$\pm$ 2.6} \\
        Subject 6 & 53.4 \rebuttal{$\pm$ 3.2} & 52.0 \rebuttal{$\pm$ 3.8} & 50.1 \rebuttal{$\pm$ 4.3} \\
        Subject 7 & 75.7 \rebuttal{$\pm$ 6.9} & 76.8 \rebuttal{$\pm$ 5.2} & 75.2 \rebuttal{$\pm$ 5.0} \\
        Subject 8 & 77.4 \rebuttal{$\pm$ 4.2} & 80.0 \rebuttal{$\pm$ 2.1} & 81.2 \rebuttal{$\pm$ 1.9} \\
        Subject 9 & 76.7 \rebuttal{$\pm$ 4.3} & 79.3 \rebuttal{$\pm$ 3.1} & 79.7 \rebuttal{$\pm$ 2.9} \\
      \cmidrule(r){1-1}
      \cmidrule(r){2-3}
      \cmidrule(r){4-4}
      \textbf{Mean}&\textbf{71.3} \rebuttal{\textbf{ $\pm$ 1.3}}& \textbf{71.2} \rebuttal{\textbf{$\pm$ 1.4}}& \textbf{70.9} \rebuttal{\textbf{$\pm$ 1.3}}\\
      Std. dev. sub$^*$ & 11.5 & 14.0 & 14.3 \\
      \toprule
    \end{tabular}
\begin{tablenotes}\footnotesize
\item $^*$Standard deviation across average accuracies per subject. 
\end{tablenotes}
\end{threeparttable}
  }
   \vspace{-.3cm}
\end{table}

\rebuttal{To obtain results comparable to literature, we strictly follow the rules of BCI Competition IV-2a for splitting the dataset, as explained in Section~\ref{ch:background:dataset}.} Table~\ref{tab:results:accuracy} compares the classification accuracy of the original \eegnet{}, the adapted \eegnet{} using ReLU activations, and the quantized Q-\eegnet{}.
\rebuttal{The training and testing procedures are implemented in PyTorch and are repeated 50 times for each subject to determine the variance in accuracy among different runs.
Table~\ref{tab:results:accuracy} reports the average accuracy and the standard deviation over the runs.
}
The network modifications (i.e., using ReLU instead of ELU) had an negligible impact of $-0.1$\% on the classification accuracy compared to the original \eegnet{}. 
Moreover, the quantization to 8-bit fixed-point yields a negligible accuracy loss of 0.3\%.

\begin{table}
  \centering
  \caption{Optimizations for Q-\eegnet{} on \wolf{} at 50\,MHz. Letters A--H refer to sections~\ref{ch:implementation:simd}--\ref{ch:eegnet:implementation:duplication}.}\label{tab:results:eegnet}
   \vspace{-.2cm}
  \resizebox{\columnwidth}{!}{%
    \footnotesize
    \begin{tabular}{lrrrrrr}
      \toprule
      &  baseline & A+B & C & D & E+F+G & H\\
      \cmidrule(r){1-1} \cmidrule(r){2-7}
      Temp.\,Conv.\,ms] & 1653.82 & 331.37 & 42.67 & 40.50 & --- & --- \\
      Spat.\,Conv.\,[ms] & 58.00 & 42.31 & 6.74 & 6.74 & --- & --- \\
      Temp.+Spat.\,[ms] & --- & --- & --- & --- & 31.08 & 26.16 \\
      Sep.\,Conv.\,[ms] & 11.90 & 6.88 & 0.94 & 0.94 & 0.81 & 0.81 \\
      FC [ms] & 0.07 & 0.07 & 0.07 & 0.07 & 0.07 & 0.07 \\
      \cmidrule(r){1-1} \cmidrule(r){2-7}
      Complete [ms] & 1732.01 & 380.27 & 50.33 & 48.04 & 31.98 & 27.06 \\
       \cmidrule(r){1-1} \cmidrule(r){2-7}
      Speedup & --- & 4.55$\times$ & 7.56$\times$ & 1.05$\times$ & 1.50$\times$ & 1.18$\times$\\
      Tot. speedup & --- & 4.55$\times$ & 34.4$\times$ & 36.1$\times$ & 54.2$\times$ & 64.0$\times$\\
      Memory [kB] & 230.29 & 248.87 & 248.87 & 248.87 & 35.41 & 68.15\\
      \glspl{mac}/cycle & 0.15 & 0.69 & 5.18 & 5.42 & 8.15 & 9.63 \\
      insn/cycle & 0.79 & 0.69 & 0.66 & 0.65 & 0.89 & 0.83 \\
      \bottomrule
    \end{tabular}
  }
  \vspace{-.3cm}
  %\begin{minipage}{\columnwidth}
    %\footnotesize
    %\begin{itemize}
    %\item[$^\ddag$] Concurrent implementation, computing the cross-correlation instead of the convolution and flipping the input feature map of block 2 and 4.
    %\item[$^\star$] The number of \glspl{mac} per cycle for the implementation with duplicated feature maps, including \gls{dma} transfers.
    %\end{itemize}
  %\end{minipage}
\end{table}

Table~\ref{tab:results:eegnet} shows the performance improvements for the optimizations on Q-\eegnet{} presented in Section~\ref{ch:eegnet:implementation}, and compares the computation time of the different parts on \wolf{}, executed at 50\,MHz.
One can notice that the execution time of the complete inference, executing all the layers at once, does not correspond precisely to the sum of each layer. This is due to the variability introduced by the measurement framework; however, the difference is negligible.
%For measuring the contribution of each layer, the compilation and execution of every separate layer was conducted isolated. 
%Therefore, the execution of the complete inference slightly deviates from the sum of the contribution of all isolated layers. 
%
%The numbers of \glspl{mac} per cycle are derived theoretically from the number of \glspl{mac} based on the network description. % and not the implementation.
%The column showing the instructions per cycle highlights the utilization of each core, giving an estimate of the number of processor stalls.
From the table, we can see that with PULP-DSP library, the execution is accelerated by 4.55$\times$ using SIMD, loop unrolling, and transposing feature maps (A\,+\,B).
Furthermore, the 7.56$\times$ speedup demonstrates that Q-\eegnet{} can be parallelized very well over eight cores using concurrent execution (C). 
With the substitution of cross-correlation instead of convolution (D), we gain another 5\% speedup. 
When combining our novel optimizations (E--H), the speedup is additionally improved by 78\%, resulting in an overall speedup of 64$\times$ with respect to the baseline implementation, highlighting the effectiveness of our proposed optimizations.
Looking more closely, the interleaved computation of the Temporal and Spatial Convolution (E) reduces the memory footprint of Q-\eegnet{} by almost 85\% from 230.29\,kB to 35.41\,kB, making it applicable to embedded devices with very limited memory availability.
Moreover, merging (F) and reordering (G) the \gls{bn} layers
%The two optimizations from Sections~\ref{ch:implementation:collect_bn}~and~\ref{ch:implementation:reordering} combined 
reduces the total number of divisions by 98\%.
This fact also explains the higher number of instructions per cycle.
Finally, the replication of the feature maps (H) eliminates the re-shuffling instructions and gives the highest performance of 9.63\,MACs/cycle. 
However, it introduces more memory usage, but still 70\% less than the baseline implementation. 
It is left then to the discretion of the user whether to include this last optimization step, depending on the available resources.

For comparison, we implement the most compute-intensive block of Q-\eegnet{}---the Temporal Convolution block contributing over 95\% to the overall \gls{mac} operations---on \wolf{} using the \textsc{PULP-NN} library~\cite{garofalo2020pulp}. 
%
%The number of \gls{mac} operations in the Temporal convolution block makes up more than 95\% of the entire network. 
%
%Hence, we implement the first temporal convolution and the subsequent batch normalization using \textsc{PULP-NN} library. 
%
The measured result shows that it
%
%The measurements do not include the \gls{dma} transfers and ignore the padding.
%
%Nevertheless, i
takes 76.13\,ms to complete, achieving only 3.28\,\glspl{mac} per cycle.
This is 3$\times$ slower than our implementation, which at the same time, also computes all remaining layers and includes all \gls{dma} transfers. 
%the second spatial convolution, batch normalization, and pooling layer, and includes all the necessary \gls{dma} transfers.
The reason can be mostly attributed to the focused optimization of PULP-NN on 2D convolutions instead of 1D convolutions, which are often encountered in \glspl{cnn} for image classification.

%\subsection{Power Measurements}
Finally, we perform power measurements to assess the energy consumption of our implementation. 
Table~\ref{tab:results:summary} shows the measurements on \wolf{}, including the startup time of the compute cluster, in two different configurations: \mbox{50\,MHz\,@\,0.8\,V} and \mbox{350\,MHz\,@\,1.2\,V}. 
The former consumes the least amount of energy per inference (0.337\,mJ) at highest energy efficiency (38.99\,GMAC/s/W), while the latter provides the highest performance (2258\,MMAC/s) by executing one inference in only 5.82\,ms at energy efficiency of 20.97\,GMAC/s/W. 
Comparing to~\cite{Wang2020AnComputing}, which has also implemented \eegnet{} at maximum clock frequency of 216\,MHz on an ARM Cortex-M7 (STM32F756ZG) using \textsc{Cube.AI}, our implementation is approximately 256$\times$ more energy-efficient.

\begin{table}
  \centering
  \caption{Comparison between Q-\eegnet{} on \wolf{} and \eegnet{} on ARM Cortex M7.}\label{tab:results:summary}
  \vspace{-.2cm}
  {
    \footnotesize
    \begin{tabular}{lrrr}
      \toprule
      Platform & \multicolumn{2}{c}{\wolf{} (ours)} & Cortex M7~\cite{Wang2020AnComputing} \\
     \cmidrule(r){1-1} \cmidrule(r){2-3} \cmidrule(r){4-4}
      Input size & \multicolumn{2}{c}{$22\times1125$} & $38\times80$ \\
      \glspl{mac} & \multicolumn{2}{c}{12\,984\,432} & 1\,509\,220 \\
      Memory & \multicolumn{2}{c}{68.15\,kB} & 146.32\,kB \\
       & @50\,MHz & @350\,MHz & @216\,MHz \\
      \cmidrule(r){1-1} \cmidrule(r){2-2}\cmidrule(r){3-3} \cmidrule(r){4-4}
      %Voltage [V]           & 0.8 & 1.2& ??\\
      Power [mW] & 11.75& 107.87 & 413.06 \\
      Time/inference [ms]    &28.67& 5.82 & 43.81 \\
      Energy/inference [mJ]   &0.337 & 0.627 & 18.1\\
      Throughput [MMAC/s] & 458 & 2258 & 34 \\
      En. eff. [GMAC/s/W] & 38.990 & 20.957 & 0.082 \\
      \toprule
    \end{tabular}
  }
  \vspace{-.3cm}
\end{table}

% How I determined the memory usage:
% Summary: 
% * Input and all feature maps remaining after interleaved computation
% * All weights (and all BN coefficients)
% * Memory required for interleaved computation (reused memory)
% * Memory required for data replication
%
% Idea: Memory that would be necessary if our optimizations would be implemented on a system which has only 1 explicit memory location (maybe some automatically managed caches).
% Reason: Counting everything twice since we move it to L1 for faster computation is imho not fair.
% Detailed:
%   22 * 1128 (Input, aligned)
% + 16 * 1 * 140 (Output spatial convolution)
% + 16 * 1 * 140 (Output of first conv in Separable conv)
% + 16 * 1 * 20  (Output of second conv in Separable conv, aligned)
% + 4            (Network output)
% + 4 * 4 * 22 * 8 (Memory for interleaved computation)
% + 6 * 22 * 248 (Memory required for data replication: 6 because 3 replications at 2 places)
% + 8 * 64 (Temporal conv weights)
% + 22 * 16 (Spatial conv weights)
% + 16 * 16 + 16 * 16 (Separable conv weights)
% + 1280 + 4 (FC weights + bias)
% + 4 * 2 * (8+16+16) (BN coefficients)
% = 68148 B

\section{Conclusion}\label{ch:conclusion}
This paper presents Q-\eegnet{}, a modified and 8-bit quantized \eegnet{}, which enables energy-efficient inference on resource-limited low-power edge devices at negligible accuracy loss. 
With the proposed optimizations, which can be adopted by other similar \gls{cnn} architectures, we achieve a runtime speedup of up to 64$\times$ relative to the baseline implementation on \wolf{}, yielding only 5.82\,ms and 0.627\,mJ per inference.
Due to its specialization, our implementation surpasses the energy-efficiency of general \gls{cnn} libraries, like \textsc{PULP-NN} and \textsc{Cube.AI}.
This work shows that MI-\gls{bmi} can be operated directly on the edge, exclusively using fixed-point operations, on a low-power embedded platform.
\rebuttal{In the future, the proposed technique of interleaved layers can be included in  automatic code generators/compilers for deep learning inference, such as Apache TVM~\cite{Apachetvm2018}, Google MLIR~\cite{GoogleMLIR}, or DORY \cite{burrello2019work}, to overcome the layer-by-layer implementation paradigm. Moreover, even lower bit representations and mixed-precision inference can be further explored.
%to further reduce the memory footprint and increase the speedup.
}

% \section*{Acknowledgment}

% The preferred spelling of the word ``acknowledgment'' in America is without 
% an ``e'' after the ``g''. Avoid the stilted expression ``one of us (R. B. 
% G.) thanks $\ldots$''. Instead, try ``R. B. G. thanks$\ldots$''. Put sponsor 
% acknowledgments in the unnumbered footnote on the first page.

\bibliographystyle{IEEEtran}
\bibliography{bibliography,ref_michael_mendeley}

\end{document}